\documentstyle[aps,epsfig]{revtex}

\begin{document}

\draft 
\title{Intermediate-mass dilepton spectra  
and the role of secondary hadronic processes in heavy-ion collisions}

\author{G.Q. Li}
\address{Department of Physics and Astronomy, State University of New
York at Stony Brook,\\
Stony Brook, New York 11794}
\author{C. Gale}
\address{Physics Department, McGill University,\\ 3600 University St., 
Montr\'eal, 
QC, H3A 2T8, Canada}

\maketitle
  
\begin{abstract}
We carry out a study of intermediate-mass (between
1 and 2.5 GeV) dilepton
spectra from hadronic interactions in heavy-ion collisions.
The processes considered are $\pi\pi\rightarrow l{\bar l}$,
$\pi\rho\rightarrow l{\bar l}$, $\pi a_1\rightarrow l{\bar l}$,
$\pi\omega\rightarrow l{\bar l}$, $K{\bar K}\rightarrow l{\bar l}$,
and $K{\bar K^*}+c.c \rightarrow l{\bar l}$. The elementary 
cross sections for those are obtained from  
chiral Lagrangians involving pseudoscalar, vector,
and axial-vector mesons. The respective electromagnetic form
factors are determined by fitting to experimental
data for the reverse processes of $e^+e^-\rightarrow hadrons$.
Based on this input we calculate cross sections and 
thermal dilepton
emission rates and compare our results with those from other
approaches. Finally we use  these elementary cross sections
with a relativistic transport model and calculate dilepton
spectra in S+W collisions at SPS energies. The comparison
of our results with experimental data from the HELIOS-3
collaboration indicates the importance of the secondary
hadronic contributions to the intermediate-mass dilepton
spectra.
 
\end{abstract}

\pacs{25.75.Dw, 12.38.Mh, 24.10.Lx}

\section{introduction}

The experimental measurement and theoretical investigation of
dilepton production constitute one
of the most active and exciting fields in the physics of 
relativistic nuclear
collisions \cite{qm96}.
Because of their relatively weak final-state interactions
with the hadronic environment, dileptons, as well as photons, are
considered ideal probes of the early stage of heavy-ion
collisions, where quark-gluon-plasma (QGP) formation is
expected \cite{shur80,kkmm86}.   Because of an additional 
variable, the invariant mass $M_{l{\bar l}}$, dileptons
have the advantage of a better signal to background
ratio than real photons \cite{itz95}. They of course also prove 
superior  in processes involving two-body annihilations. 

Dilepton mass spectra produced in heavy ion collisions can 
basically be divided into three regions.
The low-mass region below $m_\phi$ ($\sim$ 1 GeV)
is dominated by hadronic interactions and hadronic
decays. In the intermediate-mass region
between $m_\phi$ and about 2.5 GeV, the contribution
from the thermalized QGP might be seen \cite{shur78,mt85,kap91}. 
In the high-mass region at
and above $m_{J/\Psi}$ the major effort in heavy ion experiments has been the
detection and understanding of $J/\Psi$ suppression.

So far, the experimental measurement of dilepton spectra 
in ultrarelativistic heavy-ion collisions has mainly been carried out
at the CERN SPS by three collaborations: the CERES
collaboration is dedicated to dielectron measurements in the 
low-mass region \cite{ceres,drees96}, the HELIOS-3 \cite{helios} 
collaboration has measured dimuon spectra from 
threshold up to the $J/\Psi$ region, and the NA38/NA50 \cite{na38} 
collaboration measures dimuon spectra in the intermediate- 
and high-mass regions, emphasizing $J/\Psi$ suppression
(for a summary of low- and intermediate-mass dilepton
measurements see Refs. \cite{drees96,tse98,wurm97}).
In addition, dilepton spectra in heavy-ion collisions at energies of a
few GeV/nucleon were measured by the DLS collaboration 
\cite{dls}.
In the near future, dilepton spectra will be measured
by the PHENIX collaboration \cite{phenix} at RHIC, 
and by the HADES collaboration at the GSI \cite{hades}.

Recent observation of the enhancement of low-mass dileptons in
central heavy-ion collisions by the CERES \cite{ceres,drees96} and
the HELIOS-3 \cite{helios} collaborations has generated a 
great deal of theoretical activity. 
Different models have been used to interpret 
these data. The results from many groups with 
standard scenarios (i.e., using vacuum meson properties) are 
in remarkable agreement with each other, but in significant 
disagreement with the data: the experimental spectra in the 
mass region from 0.3-0.6 GeV are substantially underestimated 
\cite{likob,others} (see also Ref. \cite{drees96}). 
This has led to the suggestion of 
various medium effects that might be responsible for the
observed enhancement. In particular, the dropping vector 
meson mass scenario \cite{likob,brown91,hat92} 
is found to provide a unified  
description of both the CERES and HELIOS-3 data. However, see also
Ref. \cite{ralf}. 

In the high-mass region around $m_{J/\Psi}$, the $J/\Psi$ suppression 
has been a subject of great interest, since it was first proposed
as a signal of the deconfinement phase transition \cite{satz86}. Various
investigations show that up to central S+Au collisions, the normal
pre-resonance absorption in nuclear matter can account 
for the observed $J/\Psi$ suppression \cite{jpsisumm}. However,
recent data from the NA50 collaboration for central Pb+Pb
collisions show an additional strong `anomalous' suppression
which might indicate the onset of the color deconfinement
\cite{na50}. 

Other interesting experimental data that have not yet received
much theoretical attention are dilepton spectra in the
intermediate-mass region from about 1 GeV to about 2.5 GeV.
Both the HELIOS-3 and NA38/NA50 collaborations have observed
significant enhancement of dilepton yield in this mass region
in central S+W and S+U collisions as compared to that in 
proton-induced reactions (normalized to the charged-particle 
multiplicity) \cite{helios,na38}. Preliminary data from the
NA50 collaboration also show significant enhancement in central
Pb+Pb collisions \cite{na38} (see also Ref. \cite{drees96}).
 
\begin{figure}[h]
\begin{center}
\epsfig{file=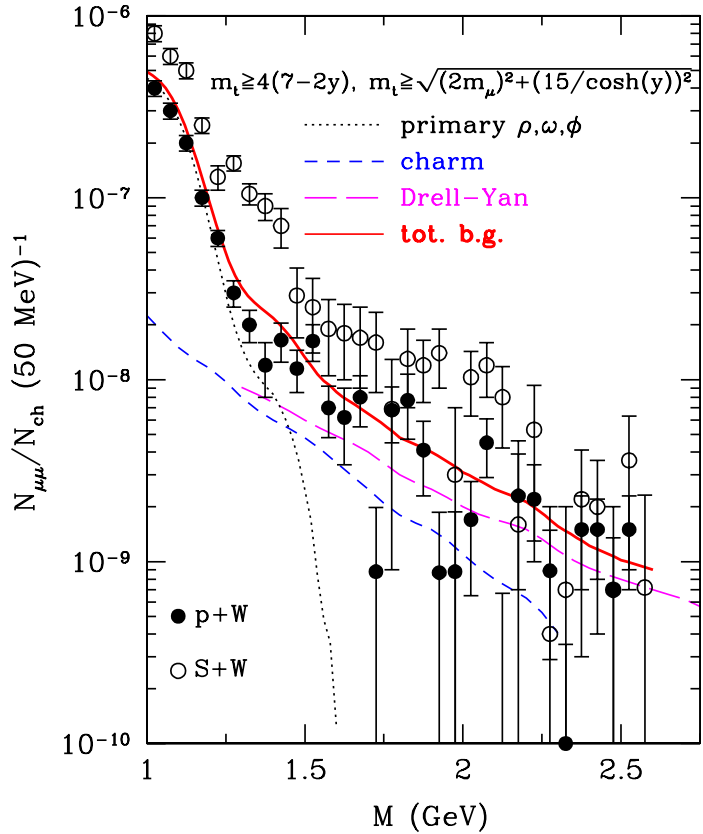,width=5in}
\caption{Comparison of background with experimental data
in p+W and S+W collisions. The data is from Ref.\protect\cite{helios}
\label{pw} }
\end{center}
\end{figure}

For dilepton spectra with mass above 1 GeV, the contributions
from charm meson decay and the initial Drell-Yan processes
begin to play a role. These hard processes 
scale almost linearly with the participant nucleon number,
and can therefore be extrapolated from proton-proton and
proton-nucleus collisions. Such a study has recently been
carried out by Braun-Munzinger {\it et al} \cite{braun97}.
The results for p+W and central S+W collisions corresponding to
the HELIOS-3 acceptance are shown in Fig. \ref{pw},
and are taken from Ref. \cite{drees96}. These, together
with the dileptons from the decay of primary vector mesons,
are collectively termed `background' in this work. It is seen that this
background describes very well the dimuon spectra in 
p+W reactions, shown in the figure by solid circles.

However, as can be from the figure, the sum of these background
sources grossly underestimates the dimuon yield in central
S+W collisions, shown in the figure by open circles.
Since the dimuon spectra are normalized by the measured
charged particle multiplicity, this underestimation indicates
additional sources to dilepton production in heavy-ion
collisions. There are at least three possible sources
for this enhancement: the additional production of
charmed mesons and/or  Drell-Yan pairs, 
a QGP formed in the collisions, and 
secondary hadronic interactions. While all
these possibilities are of interest, and may actually
coexist, in this work we concentrate on the
contributions from the secondary hadronic interactions,
which we believe need to be quantitatively assessed. 
In this work we limit ourselves to meson interactions. However we will
also comment on the role of baryons later. 
For dilepton spectra
at low invariant masses, it is well known that the
$\pi\pi$ annihilation plays an extremely important role in
heavy-ion collisions. It is also expected that 
other secondary processes will play a role in the
dilepton spectra in the intermediate mass region, and we attempt 
to demonstrate this in a realistic 
calculation that compares with experimental data.
%
%
This will be done with  
the relativistic transport 
model used in Ref. \cite{likob,lib97} for the study of low-mass 
dilepton and
photon production. The main motivation for such a study is to
understand the origin of the observed enhancement in the 
intermediate-mass region, and to see whether the data calls 
for the formation of a QGP phase. 

In Section 2 we discuss the elementary dilepton production
cross sections that are needed as inputs in the transport 
model. We shall emphasize the constraints imposed on these
cross sections by the experimental data for the reverse 
process of $e^+e^-\rightarrow hadrons$. In Section 3, we
discuss the thermal dilepton emission rates from the 
hadronic interactions, and compare our results with those from
other approaches. The elementary cross sections, as constrainted
by the $e^+e^-$ annihilation data, are then used in the transport
model to calculate dilepton spectra in heavy-ion collisions
at SPS energies. The results will be presented in Section 4.
The paper ends with a brief summary and outlook in 
Section 5.
 
\section{cross sections and form factors}

In the low-mass region from 2$m_\pi$ to $m_{\rho^0,\omega}$,
it has been shown that $\pi^+\pi^-$ annihilation, which is
characteristic of heavy-ion collisions, plays  an  important role
in explaining the observed enhancement. Similarly, it
is expected that other secondary hadronic processes
should also play some role in the intermediate-mass region.
Indeed, previous thermal rate calculations based on kinetic 
theory show that in the mass and temperature regions relevant for
this study, the following processes (from the hadronic
phase) are important: $\pi\pi\rightarrow l{\bar l}$,
$\pi\rho\rightarrow l{\bar l}$, $\pi a_1\rightarrow l{\bar l}$,
$\pi\omega\rightarrow l{\bar l}$, $K{\bar K}\rightarrow l{\bar l}$,
and $K{\bar K^*}+c.c \rightarrow l{\bar l}$ 
\cite{gale94,song94,haglin95,kim96}. 
Among these processes, $\pi a_1\rightarrow l{\bar l}$ has 
been found to be the most important one, mainly because of its
large cross section \cite{song94,kim96} (similar conclusions
have been drawn for thermal photon production \cite{xiong92,song93}).
Note here that we discuss only two-body reactions, as they 
should dominate the phase space region we are considering.

An important input in the transport model calculation of dilepton
spectra in heavy-ion collisions is the elementary dilepton
production cross sections for the processes outlined above.
In next three subsections, we will discuss the cross sections
and form factors for pseudoscalar-pseudoscalar, pseudoscalar-vector,
and pseudoscalar-axial vector meson annihilation.

\subsection{pseudoscalar-pseudoscalar annihilation}

\vskip 1cm
\begin{figure}[htp]
\begin{center}
\epsfig{file=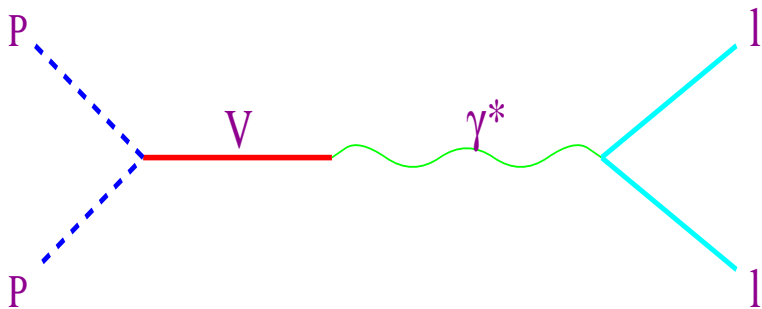,height=4cm,width=10cm}
\caption{Feynman diagram for the annihilation of two
pseudoscalar mesons into a lepton pair in the VMD model. \label{ppfeyn}}
\end{center}
\end{figure}

In this class we have $\pi-\pi$ and $K-{\bar K}$ annihilation.
In the Vector Meson Dominance model (VMD), 
the Feynman diagram for this process is quite simple and is shown 
in Fig. \ref{ppfeyn},
where $P$ represents a pseudoscalar meson, $V$ the intermediate
vector meson, $\gamma ^*$ the virtual photon, and $l{\bar l}$
the lepton pair. 
See Ref. \cite{ben98} for a discussion of the possible uncertainties in
VMD. 
We use a standard Lagrangian
for the pseudoscalar-pseudoscalar-vector interaction,
\begin{eqnarray}
{\cal L}_{VPP} = g_{VPP} V^\mu P \stackrel{\leftrightarrow}{\partial}_\mu P  
\end{eqnarray}

The obtained dilepton production cross section in pion-pion annihilation
is well known \cite{gale87,liko95} to be
\begin{eqnarray}\label{ppee}
\sigma (\pi^+\pi^-\rightarrow l{\bar l})
={8\pi \alpha^2 k \over 3M^3} |F_\pi (M)|^2 (1-{4m_l^2\over M^2})
(1+{2m_l^2\over M^2}),
\end{eqnarray}
where $k$ is the magnitude of the three-momentum of the pion
in the center-of-mass frame, $M$ is the mass of the
lepton pair, 
and $m_l$ is the mass of the lepton. It is well
known that the electromagnetic form factors $|F_\pi (M)|^2$
play important role in this process, providing empirical support for 
VMD: the pion electromagnetic
form factor is dominated by the $\rho (770)$ meson.
In addition, at large invariant masses, higher $\rho$-like 
resonances such as $\rho (1450)$ were found to be important \cite{bia91}.

\begin{figure}[hbt]
\begin{center}
\epsfig{file=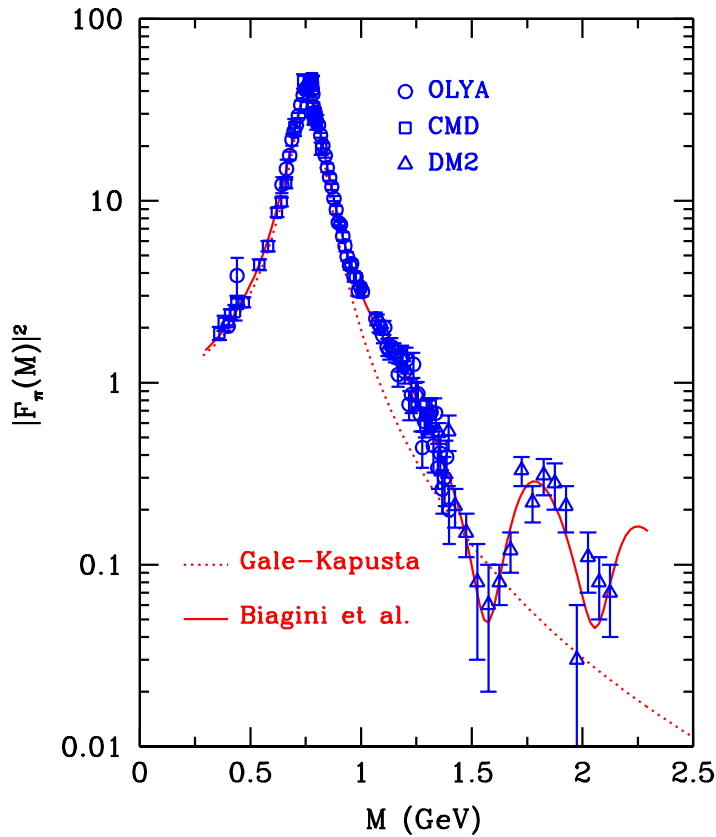,height=5in,width=5in}
\caption{The pion electromagnetic form factor. The solid
curve is based on the model of \protect\cite{bia91},
while the dotted curve is based on the parameterization
of one of us with Kapusta \protect\cite{gale87}. The experimental
data are from the OLYA collaboration \protect\cite{olyapi} (circles), 
the CMD collaboration \protect\cite{olyapi} (squares), 
and the DM2 collaboration \protect\cite{dm2pi} (triangles).
\label{ppform}}
\end{center}
\end{figure}

\begin{figure}[hbt]
\begin{center}
\epsfig{file=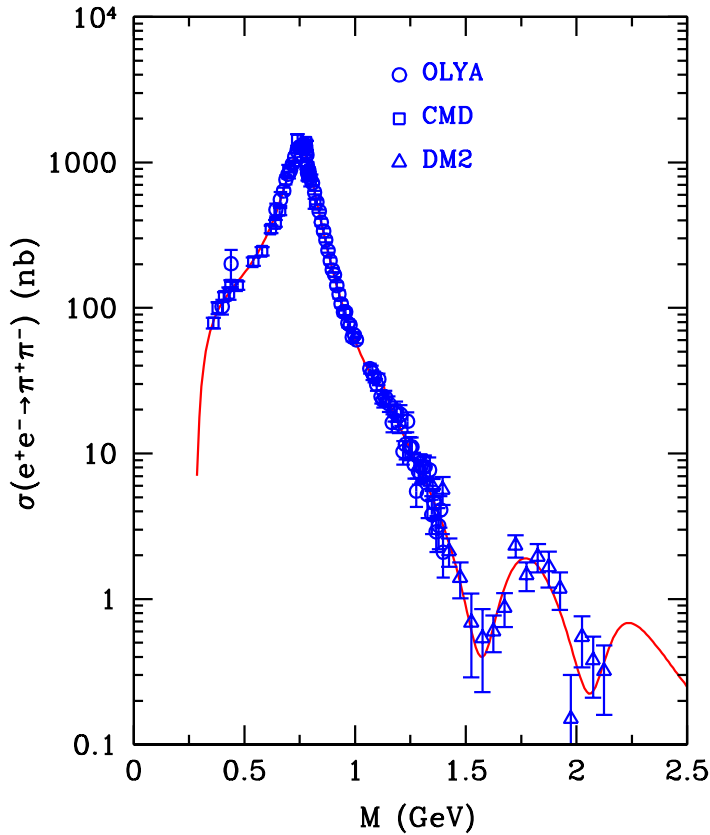,height=5in,width=5in}
\caption{The cross section for $e^+e^- \rightarrow 
\pi^+\pi^-$. The solid curve is based on the model of 
\protect\cite{bia91}. The experimental data are from 
the OLYA collaboration \protect\cite{olyapi} (circles), 
the CMD collaboration \protect\cite{olyapi} (squares), 
and the DM2 collaboration \protect\cite{dm2pi} (triangles).
\label{ppxsect}}
\end{center}
\end{figure}

Using Eq. (\ref{ppee}) and detailed-balance, we
can get the cross section for $\pi^+\pi^-$ production in 
$e^+e^-$ annihilation,
\begin{eqnarray}
\sigma _(e^+e^-\rightarrow \pi^+\pi^-) 
={8\pi \alpha^2 k^3 \over 3M^5} |F_\pi (M)|^2 
\end{eqnarray}
This cross section has been measured with high accuracy 
in the mass region that is relevant to this study \cite{olyapi,dm2pi}. In 
Ref. \cite{bia91} a detailed analysis of the experimental
data was carried in order to determine the pion 
electromagnetic form factor. Four $\rho$-like vector mesons  
were found to be present. The comparison of the form factor
determined in Ref. \cite{bia91} with the experimental data
from the OLYA collaboration \cite{olyapi} (circles), 
the CMD collaboration \cite{olyapi} (squares), 
and the DM2 collaboration \cite{dm2pi} (triangles)
is shown in Fig. \ref{ppform}. A 
direct comparison with the experimental cross section
is shown in Fig. \ref{ppxsect}. The agreement between the
model and the experimental data is excellent. This assures
us that the elementary dilepton production cross section 
in pion-pion annihilation used in our transport model is
well under control.

The pion electromagnetic form factor has been parameterized in
different ways. For example, in Ref. \cite{gale87},
one of us and  Kapusta proposed the following form,
\begin{eqnarray}\label{gk87}
|F_\pi (M)| ^2 = {m_r^4\over (M^2-m_r^{\prime 2} )^2
+(m_r\Gamma _r)^2},
\end{eqnarray}
where $m_r=0.775$ MeV, $m_r^\prime = 0.761$ MeV, and
$\Gamma _r=0.118$ MeV.
This parameterization reproduces the Gounaris-Sakurai \cite{gosa} formula and 
describes very well the experimental
form factor from the $2\pi$ threshold to about 1 GeV in
invariant mass, as shown in Fig. \ref{ppform}
by the dotted curve. It however underestimates the data
at larger invariant masses, since it neglects the role of
higher $\rho$-like resonances.

\begin{figure}[hbt]
\begin{center}
\epsfig{file=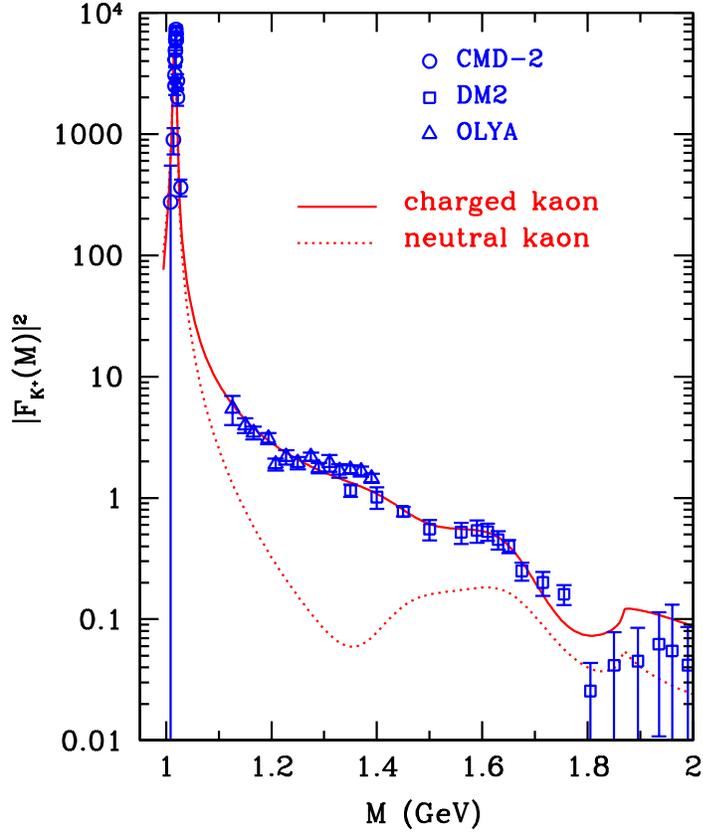,height=5in,width=5in}
\caption{The kaon electromagnetic form factor. The solid
and dotted curves are for the charged and 
neutral kaons, respectively  \protect\cite{bia91}.
The symbols are the experimental data for the charged kaon 
from the CMD-2 collaboration \protect\cite{cmd2} (circles), 
the DM2 collaboration \protect\cite{dm2kaon} (squares), and the OLYA 
collaboration \protect\cite{olyakaon} (triangles).\label{kkform}}
\end{center}
\end{figure}

\begin{figure}[htb]
\begin{center}
\epsfig{file=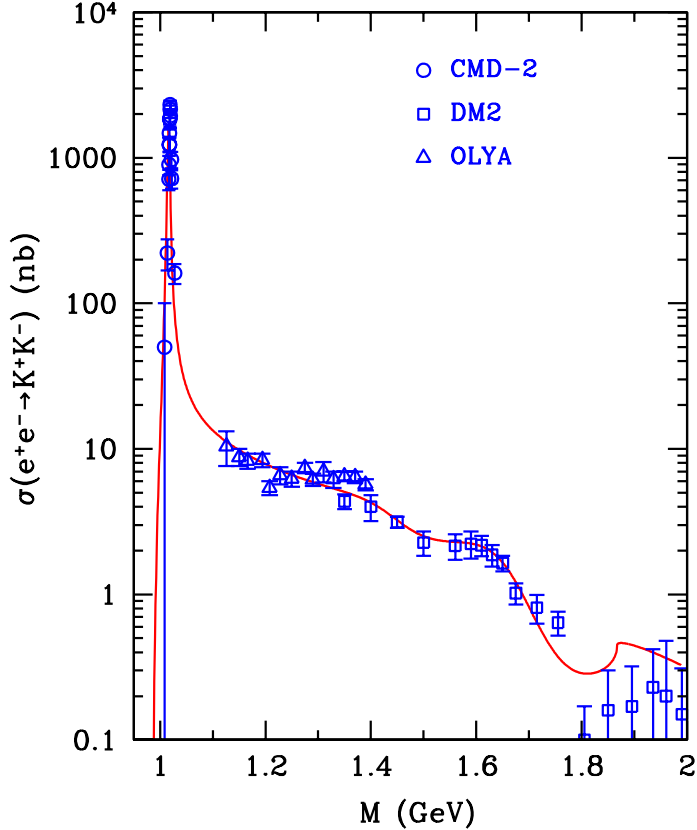,height=5in,width=5in}
\caption{The cross section for $e^+e^- \rightarrow 
K^+K^-$. The solid curve is based on the model of 
\protect\cite{bia91}. The experimental
data are from the CMD-2 collaboration \protect\cite{cmd2} (circles), 
the DM2 collaboration \protect\cite{dm2kaon} (squares), and the OLYA 
collaboration \protect\cite{olyakaon} (triangles).
\label{kkxsect}}
\end{center}
\end{figure}

The dilepton production cross section in kaon-antikaon annihilation
is very similar,
\begin{eqnarray}
\sigma (K^+K^-, K^0{\bar K}^0 \rightarrow l{\bar l})
={8\pi \alpha^2 k \over 3M^3} |F_{K^+,K^0} (M)|^2 (1-{4m_l^2\over M^2})
(1+{2m_l^2\over M^2}),
\end{eqnarray}
where $k$ is the magnitude of the kaon momentum in the
center-of-mass frame.
The cross section for the reverse process of electron-positron
annihilation can again be obtained from the
detailed-balance relation,
\begin{eqnarray}
\sigma _(e^+e^- \rightarrow K^+K^-, K^0{\bar K}^0 )
={8\pi \alpha^2 k^3 \over 3M^5} |F_{K^+,K^0} (M)|^2 
\end{eqnarray}
In these equations, $|F_{K^+}|^2$ and $|F_{K^0}|^2$ are the
electromagnetic form factors of charged and neutral
kaons, respectively. These form factor are dominated by the
phi meson, $\phi (1020)$. At higher masses, other vector
mesons may become important \cite{bia91}.
The experimental form factor will
be used in this work. It   is shown 
in Fig. \ref{kkform}. The solid and dotted curves
are for charged and neutral kaons respectively, and
the symbols are the experimental data for the charged
kaon form factor from the CMD-2 collaboration \cite{cmd2} (circles),
the DM2 collaboration \cite{dm2kaon} (squares), and 
the OLYA collaboration \cite{olyakaon} (triangles). The neutral kaon
form factor follows from general arguments \cite{bia91}. The direct
comparison with the experimental charged kaon cross section is shown
in Fig. \ref{kkxsect}. Again, very good agreement between
the model and the data is seen.

\subsection{pseudoscalar-vector meson annihilation}

In this class we include the following processes, 
$\pi\rho\rightarrow l{\bar l}$, ${\bar K}K^*+c.c.\rightarrow
l{\bar l}$, and $\pi\omega \rightarrow l{\bar l}$.
The first two processes effectively involve three pions,
while the third one involves four pions. In transport
models a process involving three or more pions in
the initial state can only be described as a two step process
with an intermediate resonance. The empirical success of 
transport approaches gives some credibility to this scenario. 
Note however that the treatment of quantum interference remains a
possible issue in this framework \cite{lich}. 


The interaction Lagrangian for this case is 
\begin{eqnarray}
{\cal L}_{VVP} = g_{VPP}\epsilon _{\mu\nu\alpha\beta} 
\partial^\mu V^\nu \partial ^\alpha V^\beta P   \ .
\end{eqnarray}

The first two channels have been studied in 
Ref. \cite{haglin95}. The cross section for $\pi\rho$
annihilation is given by
\begin{eqnarray}
\sigma (\pi^+\rho^- \rightarrow l{\bar l})
= {2\pi \alpha ^2  k_\pi\over 9 M} |F_{\pi\rho} |^2 
\Big(1-{4m_l^2\over M^2}\Big) \Big(1+{2m_l^2\over M^2}\Big),
\end{eqnarray}
where $k_\pi$ is the magnitude of the pion momentum in
the center-of-mass frame. Note that the above cross section 
is evaluated in the narrow-width approximation  for illustration 
purposes only. This simplification is not used in the actual
transport calculation.  The cross section for $e^+e^-
\rightarrow \pi\rho$ can again be obtained from 
detailed-balance.
 
\begin{figure}[htb]
\begin{center}
\epsfig{file=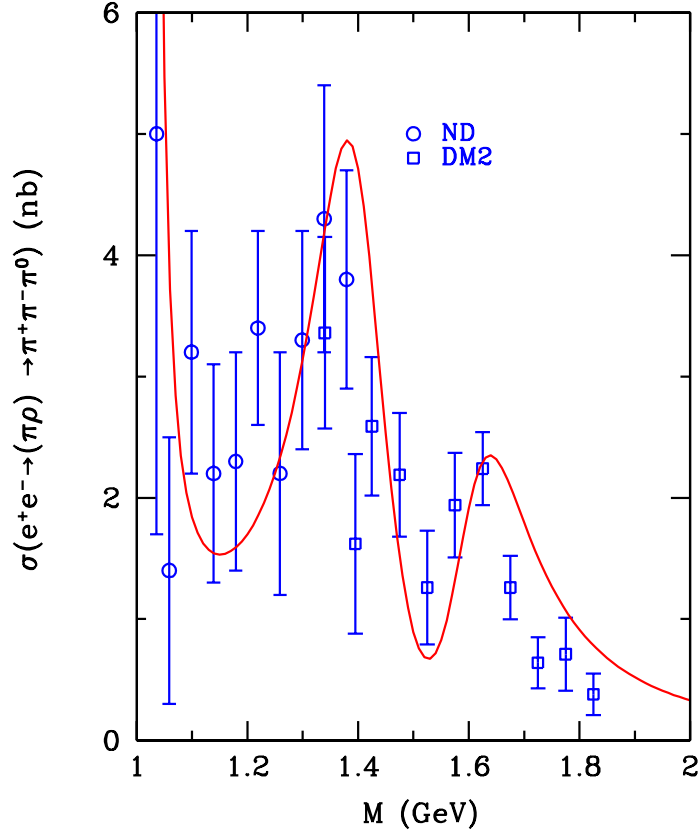,height=5in,width=5in}
\caption{The cross section for $e^+e^- \rightarrow 
(\pi\rho ) \rightarrow \pi^+\pi^-\pi^0$. 
The solid curve is our fit to  
the experimental data from the ND collaboration 
\protect\cite{nd} (circles) and the DM2 collaboration 
\protect\cite{dm2piro} (squares).
\label{piroxsect}}
\end{center}
\end{figure}

The electromagnetic form factor
$|F_{\pi\rho} (M)|^2$ can then be determined by analyzing
the experimental data for $e^+e^-\rightarrow \pi^+\pi^-\pi^0$.
In Ref. \cite{haglin95}, three isoscalar vector mesons,
$\phi (1020)$, $\omega (1420)$, and $\omega (1670)$
were found to be important in order to fit the experimental
data from Refs. \cite{aul86,bal87} (see also Ref. \cite{donn89}),
namely,
\begin{eqnarray}
F_{\pi\rho} (M) = \sum _V \left({g_{V\pi\rho}\over g_V}\right)
{e^{i\phi _V} m_V^2\over (m_V^2-M^2) - i m_V \Gamma _V}.
\end{eqnarray}
Here the summation runs over the three vector mesons listed above.
While the coupling constants $g_\phi$ and $g_{\phi\pi\rho}$
can be determined from the measured widths, the coupling
constants for other two mesons and the relative phases
were determined by fitting 
to the experimental data of Refs. \cite{aul86,bal87}. 
In this work, we determine
these coupling constants by fitting to the latest data
from the DM2 collaboration \cite{dm2piro} and the ND collaboration
\cite{nd}. These parameters are only slightly 
different from those listed in Ref. \cite{haglin95}.
The comparison of our fit with the experimental data
is  shown in Fig. \ref{piroxsect}.

\begin{figure}[htb]
\begin{center}
\epsfig{file=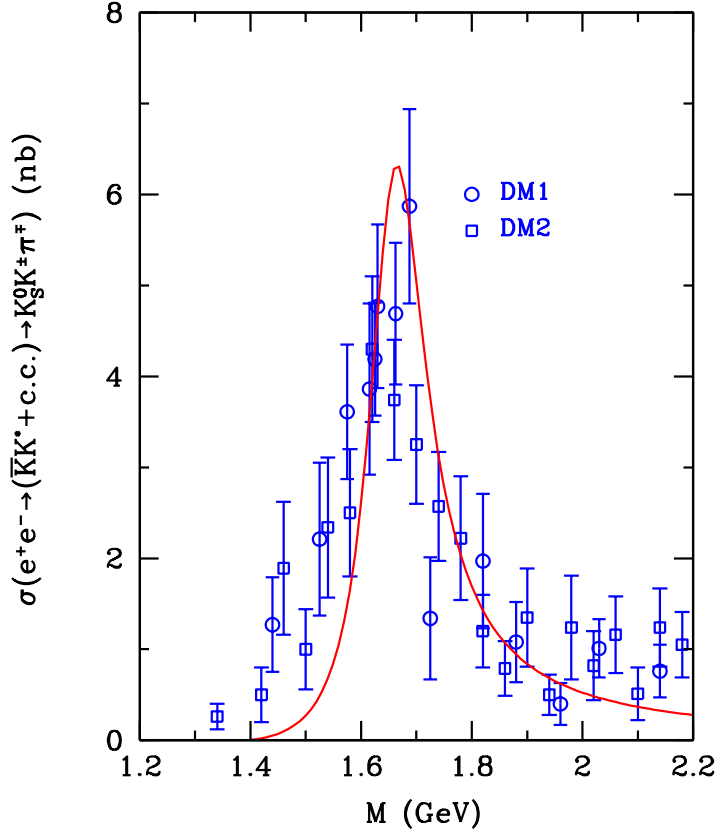,height=5in,width=5in}
\caption{The cross section for $e^+e^- \rightarrow 
(K{\bar K^*}+c.c.) \rightarrow K^0 K^\pm \pi ^\mp$. 
The solid curve is based on the model of Ref. \protect\cite{haglin95}. 
The experimental data are from the DM1 collaboration 
\protect\cite{dm1} (circles) and the DM2 
collaboration \protect\cite{dm2kks} (squares).
\label{kksxsect}}
\end{center}
\end{figure}

Similarly, the cross section for $K{\bar K^*}$ (or ${\bar K}K^*$)
is given by
\begin{eqnarray}
\sigma (K^+K^{*-}\rightarrow l{\bar l})
= {\pi \alpha ^2 k_K\over 6M} |F_{K{\bar K^*}} |^2 
\Big(1-{4m_l^2\over M^2}\Big) \Big(1+{2m_l^2\over M^2}\Big),
\end{eqnarray}
where $k_K$ is the magnitude of the kaon momentum in the
center-of-mass frame.
Since a $K^*$ eventually decays into a kaon and a pion, the
electromagnetic form factor in the above equation can
be determined by analyzing the experimental data
for $e^+e^-\rightarrow K^0K^\pm \pi^\mp$,
as was done in Ref. \cite{haglin95}. The form factor was
found to be dominated by $\phi ^\prime (1680)$,
\begin{eqnarray}
F_{K{\bar K}^*}(M) = {g_{\phi^\prime K{\bar K^*}}\over
g_{\phi^\prime}} {m_{\phi^\prime}^2\over 
m_{\phi^\prime}^2 - M^2 - i m_{\phi^\prime}
\Gamma _{\phi ^\prime}},
\end{eqnarray}
where the coupling constant $ {g_{\phi^\prime K{\bar K^*}}\over
g_{\phi^\prime}} = 0.19 ~$GeV$^{-1}$ was determined by fitting
the experimental data of the DM1 collaboration \cite{dm1} and the
DM2 collaboration \cite{dm2kks}.
The comparison with the experimental data is shown in 
Fig. \ref{kksxsect}.

\begin{figure}[htb]
\begin{center}
\epsfig{file=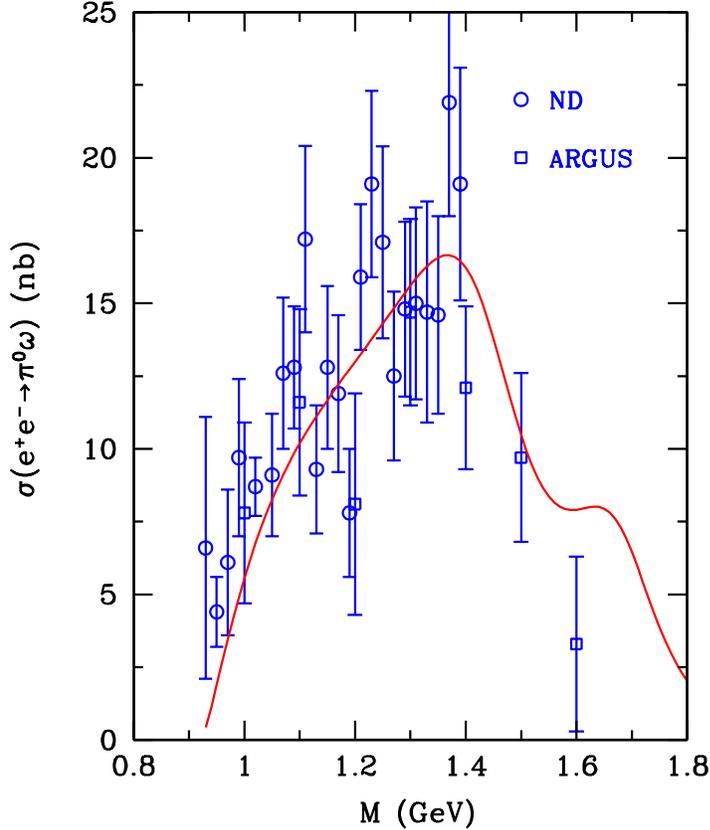,height=5in,width=5in}
\caption{The cross section for $e^+e^- \rightarrow 
\pi^0\omega$. The solid curve is based on 
the model of Ref. \protect\cite{nd}. 
The experimental data are from the ND collaboration 
\protect\cite{nd} (circles) and the ARGUS collaboration 
\protect\cite{argus} (squares).
\label{piomxsect}}
\end{center}
\end{figure}

Finally, the cross section for lepton pair production
in pion-omega annihilation is given by
\begin{eqnarray}
\sigma (\pi^0\omega \rightarrow l{\bar l})
= {4\pi \alpha ^2 k_\pi \over 9 M} |F_{\pi\omega} |^2 
\Big(1-{4m_l^2\over M^2}\Big) \Big(1+{2m_l^2\over M^2}\Big),
\end{eqnarray}
where $k_\pi$ is the magnitude of pion momentum in the
center-of-mass frame. Experimentally, the electromagnetic
form factor $|F_{\pi\omega} (M)|^2$ was studied
in Ref. \cite{nd} by measuring $e^+e^-\rightarrow \pi^0\pi^0\gamma$.
The form factor was parameterized in terms of
three isovector $\rho$-like vector mesons, $\rho (770)$,
$\rho (1450)$, and $\rho (1700)$ in Ref. \cite{nd},
\begin{eqnarray}
F_{\pi\omega} (M) = \sum _V \left({g_{V\pi\omega}\over g_V}\right)
{e^{i\phi _V} m_V^2\over (m_V^2-M^2) - i m_V \Gamma _V}.
\end{eqnarray}
Here the summation runs over the three $\rho$-like resonances
listed above. We will use the form factor determined in
Ref. \cite{nd} in this work. The comparison with the
experimental data of the ND \cite{nd}
and ARGUS collaborations \cite{argus} 
is shown in Fig. \ref{piomxsect}.

\subsection{pseudoscalar-axial vector meson annihilation}

In this class we consider chiefly $\pi a_1 \rightarrow l{\bar l}$,
which is effectively a four pion process which we treat in the manner 
described previously. 


Thermal rate calculations indicate that this process is
particularly important in the intermediate-mass region.
There are, however, significant differences between 
calculations.
Already in Ref. \cite{kim96} it was shown that the
dilepton production rates based on different models
for the $\pi\rho a_1$ dynamics can differ by an order of magnitude.
This problem was recent revisited in Ref. \cite{gao97}, where
a comparative study was carried out for both on-shell
properties and dilepton production rates using  
several models for the $\pi\rho a_1$ dynamics.
It was found that, although some of the models
provide reasonable description of the on-shell properties, the
corresponding dilepton rate could vary widely. 
By using some information from experimentally-constrained spectral
function \cite{huang95}, it was found that the effective
chiral Lagrangian of Ref. \cite{comm84}, in which the vector mesons
are introduced as massive Yang-Mills fields of the chiral
symmetry, provide reasonable off-shell as well as on-shell
properties of the $\pi\rho a_1$ system. The interaction Lagrangian
in this model is slightly elaborate and is given in 
Refs. \cite{song94,gao97}. 
The dilepton production cross section can then be obtained \cite{gao97}
\begin{eqnarray}
\sigma (\pi^+ a_1\rightarrow l{\bar l})
= {\pi \alpha ^2 {\cal H} \over 72m_{a_1}^2 g_\rho^2 M^5 k_\pi}
|F_{\pi a_1} |^2 
\Big(1-{4m_l^2\over M^2}\Big) \Big(1+{2m_l^2\over M^2}\Big)
\end{eqnarray}
where ${\cal H}$ is a complicated function of coupling constants,
masses, and kinematics, and $k_\pi$ is the magnitude
of the pion momentum in the center-of-mass frame.

\begin{figure}[htb]
\begin{center}
\epsfig{file=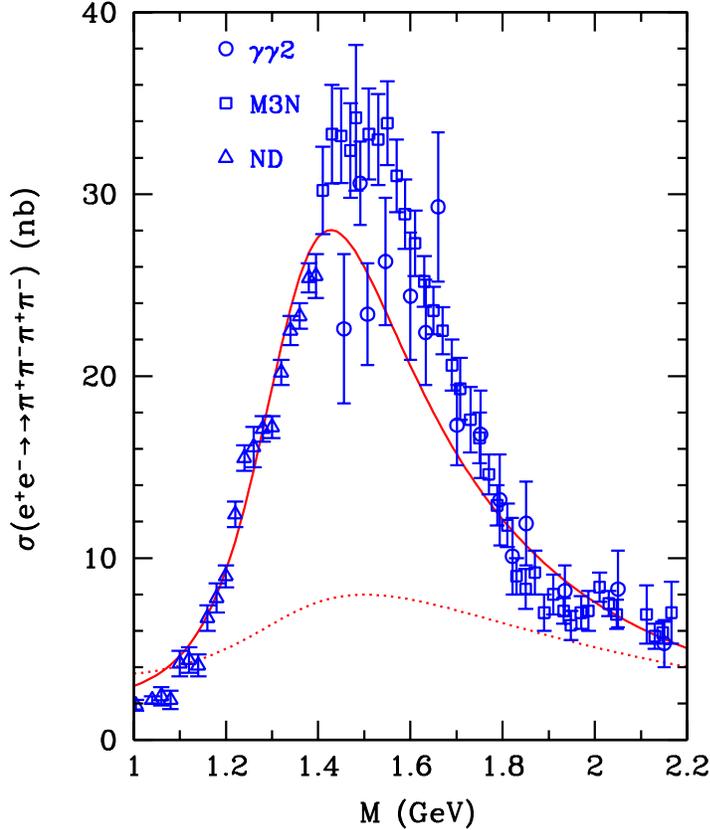,height=5in,width=5in}
\caption{The cross section for $e^+e^- \rightarrow 
\pi^+\pi^-\pi^+\pi^-$. The solid and dotted curves give
our results in the first and second scenarios for the
$\pi a_1$ form factor, respectively. 
The experimental data are from the $\gamma\gamma 2$ collaboration (circles),
\protect\cite{rr2}, the M3N collaboration \protect\cite{m3nc} 
(squares), and the ND collaboration \protect\cite{nd} (triangles).
\label{pia1xsect}}
\end{center}
\end{figure}

\begin{figure}[htb]
\begin{center}
\epsfig{file=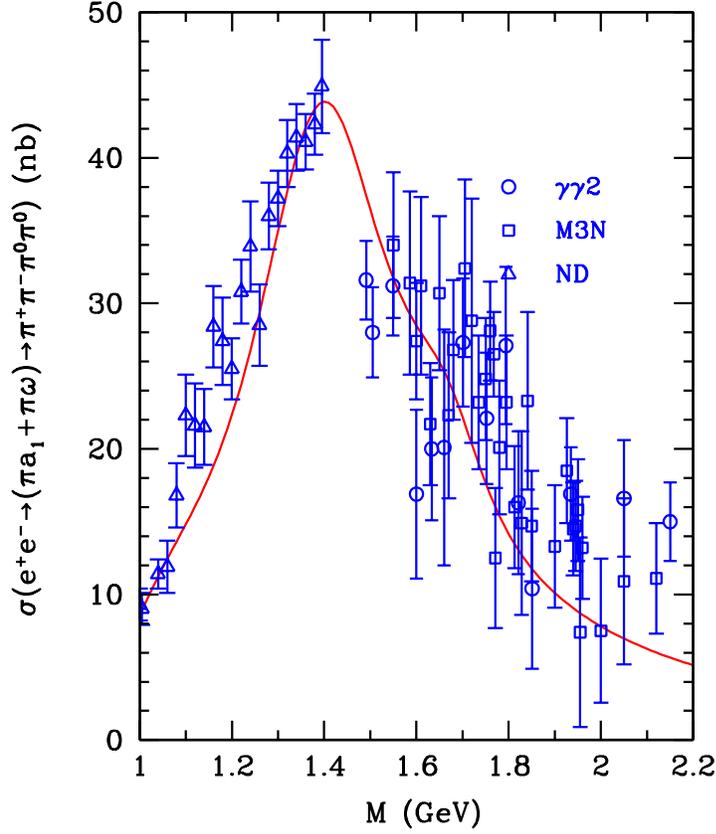,height=5in,width=5in}
\caption{The cross section for $e^+e^- \rightarrow 
\pi^+\pi^-\pi^0\pi^0$. The solid curve gives our fit 
in the first scenario for the $\pi a_1$ form factor. 
The experimental data are from the $\gamma\gamma 2$ collaboration,
\protect\cite{rr2}, the M3N collaboration \protect\cite{m3nn} 
(squares), and the ND collaboration \protect\cite{nd} (triangles).
\label{pia10xsect}}
\end{center}
\end{figure}

One must now consider the issue of the 
electromagnetic form factor, $|F_{\pi a_1}|$. In principle, it can be
determined by analyzing $e^+e^-\rightarrow \pi^+\pi^-\pi^+\pi^-$
and $e^+e^-\rightarrow \pi^+\pi^- \pi^0 \pi^0$ data. 
Although many analysis have been carried out, an
unambiguous determination of this form factor is not
yet possible, since other intermediate-state can
contribute to the same four-pion final state.
Here we consider three scenarios, which should
bracket this form factor. In the first scenario, we assume that
most of the strength in the $\pi^+\pi^-\pi^+\pi^-$ channel comes 
from a  $\pi a_1$ intermediate state. We can then determine 
$|F_{\pi a_1} (M)|^2$ from experimental
data for $e^+e^-\rightarrow \pi^+\pi^-\pi^+\pi^-$
from the $\gamma \gamma 2$  collaboration \cite{rr2}, the M3N 
collaboration \cite{m3nc}, and the ND collaboration 
\cite{nd}. Further constraint
on the $|F_{\pi a_1}|^2$ is provided by the experimental
data for $e^+e^-\rightarrow \pi^+\pi^-\pi^0\pi^0$,
which can also come from $\pi^0\omega $ intermediate state.
This scenario basically sets an upper limit for the
dilepton production from the $\pi a_1$ annihilation.
In a second scenario, we assume that the $\pi a_1$ 
electromagnetic form factor is represented by the
$\rho (770)$ only.  We shall use the the parameterization of 
Eq. (4) for this purpose. The comparison
of our results for $e^+e^-\rightarrow \pi^+\pi^-\pi^+\pi^-$
in the two scenarios with the experimental data
is shown in Fig. \ref{pia1xsect}. Our cross section
in the second scenario is similar to that extracted
in Ref. \cite{penso}. In the second scenario, the
missing strength is attributed to 
$\rho (1700)$ which then directly decays into $\rho \pi\pi$ without
coupling to the $\pi a_1$ state. In Fig.
\ref{pia10xsect}, we show the comparison of our
results in the first scenario for the process
$e^+e^-\rightarrow \pi^+\pi^-\pi^0\pi^0$.
The solid curve is the sum of our parametrizations for
$\sigma _{e^+e^- \rightarrow \pi a_1}$ and
$\sigma _{e^+e^- \rightarrow \pi \omega}$. Note that in this first
scenario, the
interpretations of the $4 \pi^{\pm}$ and $2 \pi^{\pm} 2 \pi^0$ data sets
are consistent: our fit is adjusted to provide a compromise between the
two sets of measurements. See Figs. \ref{pia1xsect} and \ref{pia10xsect}. 
 
\begin{figure}[htb]
\begin{center}
\epsfig{file=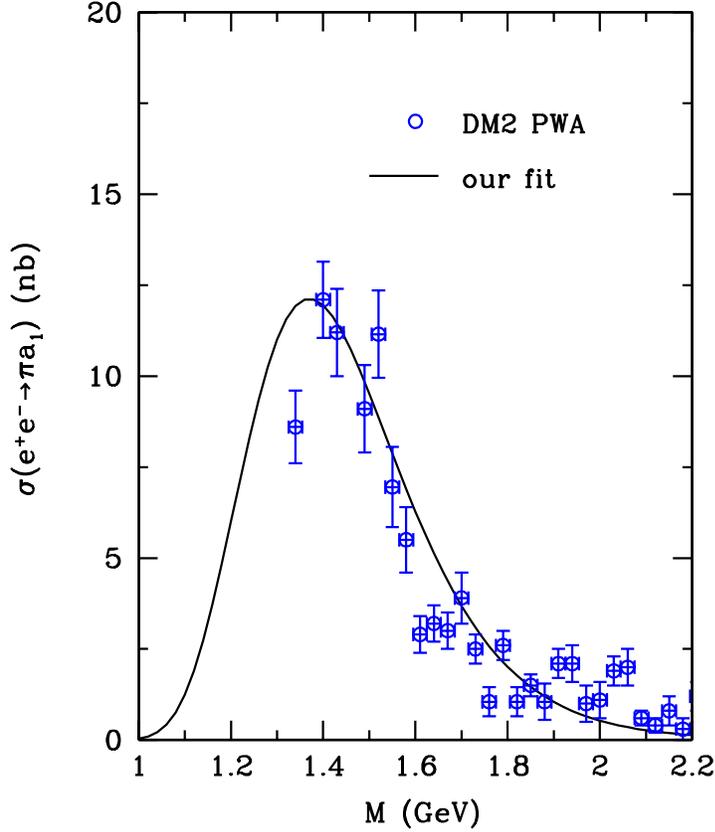,height=5in,width=5in}
\caption{The cross section for $e^+e^- \rightarrow 
\pi a_1$. The open circles are the experimental data
determined by the DM2 collaboration using 
partial wave analysis \protect\cite{dm291}.
The solid curve gives our fit to the data.
\label{eepia1dm2}}
\end{center}
\end{figure}

Finally, as a lower bound we use the cross section $\sigma _{e^+e^-
\rightarrow \pi a_1}$ determined by the DM2 collaboration
using partial wave analysis (PWA) \cite{dm291}. 
We first parameterize the data as shown in Fig. \ref{eepia1dm2},
and then use detailed-balance to obtain the
cross section $\sigma _{\pi a_1 \rightarrow l{\bar l}}$. Note that the
cross section at high masses is smaller than with our other two
approaches. 

\section{thermal rates}

The theoretical description of heavy-ion collision dynamics can 
be divided into two broad categories: transport (cascade)
models and hydrodynamical models. In the latter model, (local)
thermal and chemical equilibrium is usually assumed. 
To calculate dilepton spectra using the hydrodynamical model, one
then needs to know the thermal dilepton emission rates,
which we discuss in this section.

According to kinetic theory,
at a given temperature $T$, the differential rate per unit
time and per unit volume for a meson pair to annihilate into
a lepton pair, $a+b\rightarrow l{\bar l}$, is given by
\begin{eqnarray}
{dR\over dM^2} = {dN\over d^4xdM^2}& =&
{\cal N} \int {d^3p_a\over 2E_a (2\pi )^2} 
 {d^3p_b\over 2E_b (2\pi )^2} {d^3p_l\over 2E_l(2\pi )^2}
 {d^3p_{\bar l}\over 2E_{\bar l} (2\pi )^2}\nonumber\\
 & \times &f(E_a)f(E_b)(2\pi )^4 | {\bar {\cal M}} |^2 \delta ^4 (p_a+p_b
-p_l-p_{\bar l})\delta (M^2 - (p_l+p_{\bar l})^2),
\end{eqnarray}
where $f$ is the Bose-Einstein distributions for mesons, and
${\cal N}$ is an 
overall spin-isospin degeneracy factor.

\begin{figure}[htb]
\begin{center}
\epsfig{file=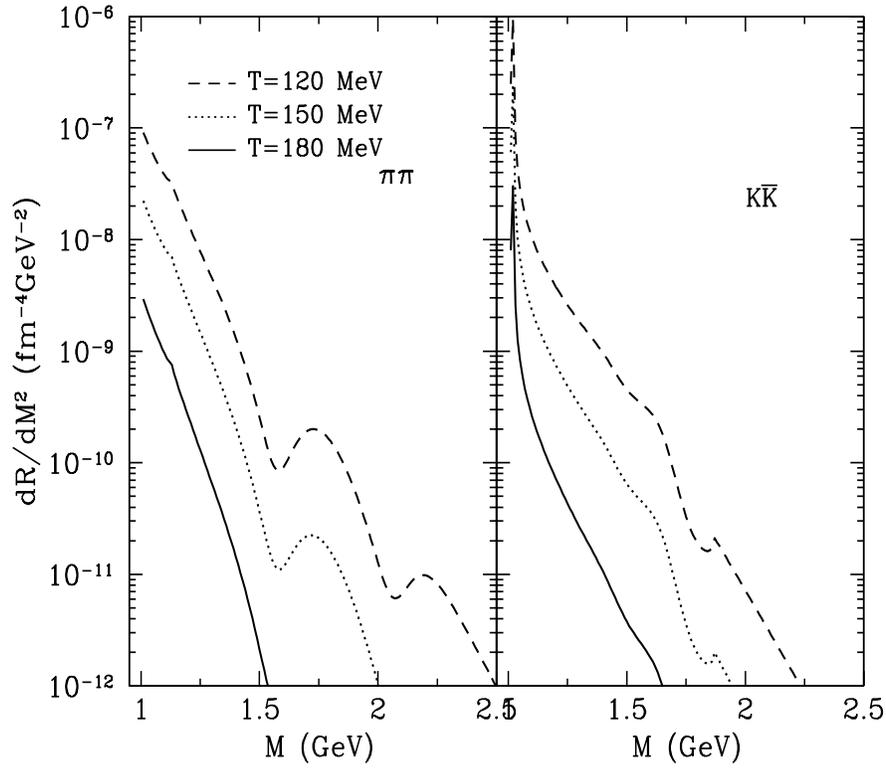,height=6in,width=5in,angle=270}
\caption{Dilepton emission rates from $\pi\pi$ and
$K{\bar K}$ annihilation at three different 
temperatures.
\label{trpk}}
\end{center}
\end{figure}

Approximating the Bose-Einstein distribution functions by
the Boltzmann ones, we arrive at a simple expression
for the thermal rate,  
\begin{eqnarray}
{dR\over dM^2} = {\cal N} {T\over 32 \pi^4 M } K_1 (M/T)
\lambda (M^2,m_a^2,m_b^2) {\bar \sigma} (a+b\rightarrow l{\bar l}),
\end{eqnarray}
where $K_1$ is a modified Bessel function,
${\bar \sigma} (a+b\rightarrow l{\bar l})$ is the isospin-averaged
cross section, and $\lambda$ is the usual kinematic
triangle relation.

\begin{figure}[htb]
\begin{center}
\epsfig{file=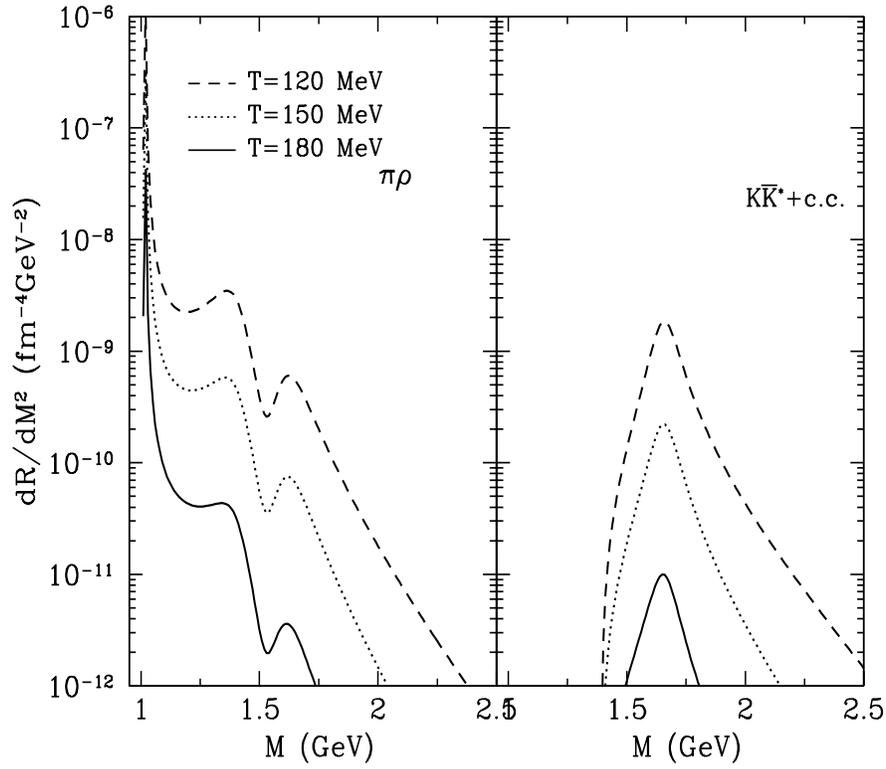,height=6in,width=5in,angle=270}
\caption{Same as Fig. \protect\ref{trpk}, for $\pi\rho$ and
$K{\bar K^*}+c.c.$ annihilation. 
\label{trpr}}
\end{center}
\end{figure}

\begin{figure}[htb]
\begin{center}
\epsfig{file=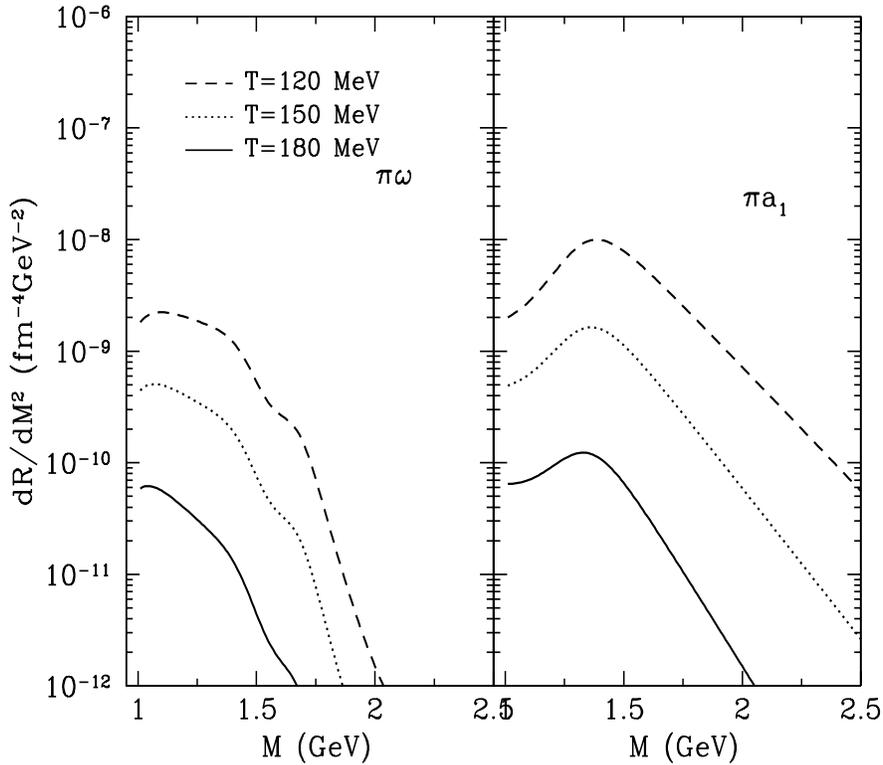,height=6in,width=5in,angle=270}
\caption{Same as Fig. \protect\ref{trpk}, for $\pi\omega$ and
$\pi a_1$ annihilation. 
\label{trpo}}
\end{center}
\end{figure}

We consider here three temperatures, $T=120$, 150, and 180 MeV.
The results for $\pi\pi$ and $K{\bar K}$ are shown in
Fig. \ref{trpk}, the results for $\pi\rho$ and $K{\bar K^*}+c.c.$
are shown in Fig. \ref{trpr}, and the results for
$\pi\omega$ and $\pi a_1$ are shown in Fig. \ref{trpo}.
The results for $\pi a_1$ are obtained in the scenario
where the measured 4$\pi$
final states all come from an intermediate  $\pi a_1$ state.
In the case of $a_1$ meson, here we have also included 
the finite-width effects as in Ref. \cite{song96}.
We mention that in the transport model, those
are included for all the baryon and meson resonances.

The dilepton emission rates are seen to depend strongly on
the temperature of the system. In heavy-ion collisions,
as the system expands, its temperature rapidly decreases,
so does the contribution to the dilepton rate. So most of
the dileptons are emitted in the early hot and dense stage
of the collisions. Comparing with previous calculations,
our results for $\pi\pi$ and $K\bar K$ are in agreement with
those of Ref. \cite{gale94}, and our results for $\pi\rho$
and $K{\bar K^*}+c.c$ are in agreement with those of Ref. \cite{haglin95}. 
Finally, as expected, our results for $\pi\omega$ and $\pi a_1$
are different from those in Refs. \cite{song94,song96},
since we use new form factors for these processes which we obtain from 
$e^+e^-$ annihilation data.
Of all the processes considered here, the $\pi a_1$ is found
to be the most important one for intermediate-mass dileptons.
This is in line with previous observations \cite{song94,kim96,song96}.

Dilepton emission rates from hot and dense hadronic matter
have been calculated in different approaches. Here we discuss
and compare our results with three of these approaches:
the spectral function approach of Ref. \cite{huang95},
the chiral reduction formalism of Refs. \cite{steele97,lee98}, and
a model based on quark-hadron duality \cite{ruus97}.

In general, the dilepton emission rate is given by the 
thermal expectation value of the electromagnetic current-current
correlation function \cite{mt85,weldon90,gaka91},
\begin{eqnarray}\label{cc}
{dR\over dM^2} = - {\alpha^2 \over 6\pi^ 3 M^2}
\int d^4x d^{-iqx} \langle\langle J^\mu (x) J_\mu (0)
\rangle\rangle_T.
\end{eqnarray}
Different approaches differ in the way this correlation function 
is approximated and calculated.
In Ref. \cite{huang95}, this current-current correlator was reduced 
to a number of spectral functions which can be 
extracted from the $e^+e^-$ annihilation
and $\tau$ lepton decay data. Including the
soft final-state interaction corrections that leads to the
parity mixing phenomenon (i.e., the mixing between the
vector and the axial-vector currents), the dilepton
emission rate can be expressed as \cite{huang95}
\begin{eqnarray}\label{huang}
{dR\over dM^2} = {2\alpha ^2\over \pi} M T K_1 (M/T)
\left( \rho ^{em} (M) - (\epsilon - {\epsilon ^2\over 2})
(\rho ^V(M) - \rho ^A (M))\right),
\end{eqnarray}
where $\epsilon = T^2/6F_\pi^2$, and $\rho^{em}$, $\rho^V$,
and $\rho^A$ are the electromagnetic spectral function, the
vector spectral function, and the axial-vector spectral
function, respectively. $\rho ^{em}$ can be related to the
total hadronic cross section of $e^+e^-$ annihilation,
\begin{eqnarray}
\rho ^{em} (M) = {M^2 \sigma (e^+e^-\rightarrow hadrons )\over
16 \pi^3 \alpha^2}.
\end{eqnarray}
Similarly, the vector spectral function can be determined by
selecting events with even number of pions
\begin{eqnarray}
\rho ^V (M)= {M^2\over 16\pi^3 \alpha^2} \sum _{n=1}
\sigma (e^+e^-\rightarrow 2n\pi ).
\end{eqnarray}
Finally, the axial-vector spectra function $\rho^A$ can
be extracted from the differential probability of the 
$\tau$ lepton decaying into an odd number of pions. Thus, with
these spectral functions extracted from the $e^+e^-$ and
$\tau$ decay data, one can determine the empirical dilepton
emission rate using Eq. (\ref{huang}). Note that the terms in $\epsilon$
represent a small correction. 

In Ref. \cite{ruus97}, the idea of a Hagedorn resonance gas
approach was explored. There, the vector meson mass
distribution is characterized by the mass spectrum $\rho _V(m)$
(we use $\rho ^V$ for the vector spectral function, and 
$\rho _V$ for the vector meson mass spectrum). The dileptons
are associated with direct vector meson decays (the
vector meson dominance assumption). Therefore the thermal
rate can expressed as
\begin{eqnarray}
{dR\over dM^2} = \rho _V (M) {\alpha ^2\over 6\pi g^2(M)}
M^2 T K_1 (M/T),
\end{eqnarray}
where $g(M)$ is the vector-meson-photon coupling constant, which also
determines the vector meson production cross section in
the $e^+e^-$ annihilation,
\begin{eqnarray}
\sigma (e^+e^-\rightarrow V) = {(2\pi)^3 \alpha^2 \over g^2(M)}
{1\over M} \rho _V(M).
\end{eqnarray}
In Ref. \cite{ruus97}, it was assumed that the total cross section
of $e^+e^-$ annihilation is saturated by the production
of vector mesons, which is supported by experimental 
observation \cite{huang95},
\begin{eqnarray}
\sigma (e^+e^-\rightarrow hadron) = R ^{exp} {4\pi \over 3} {\alpha^2
\over M^2} = \sigma (e^+e^-\rightarrow V),
\end{eqnarray}
where $R^{exp}$ is the experimental $R$ value, which reflects the
strong interaction aspect of the  $e^+e^-$ annihilation physics.
With this approximation the dilepton emission
rate can then be written as
\begin{eqnarray}
{dR\over dM^2} = R^{exp} {\alpha^2 \over 6\pi^2} M T K_1 (M/T).
\end{eqnarray}
This is very similar to the results of Ref. \cite{huang95},
except here that the soft-pion corrections
are neglected. In Ref. \cite{ruus97}, it was further assumed
that $R^{exp}$ can be approximated by $R^{part} =
N_c \sum _f e_f^2$, the parton model prediction for 
$R$. The dilepton emission rate is then determined
theoretically without referring to experimental data,
\begin{eqnarray}\label{ruus}
{dR\over dM^2} = R^{part} {\alpha^2 \over 6\pi^2} M T K_1 (M/T).
\end{eqnarray}
For three flavor case, $R^{part} = 3 (4/9 + 2/9 + 2/9) = 8/3$.
This equality states that the dilepton emission rate
from hadronic gas can be approximated by the parton model
prediction with the same accuracy as the parton model describes
the total hadronic production cross section in $e^+e^-$
annihilation. This approximation would apply to
invariant masses greater than $\approx$ 1.5 GeV. 

Finally, in Refs. \cite{steele97,lee98}, a chiral reduction
formalism \cite{zahed96} was use to reduce the 
current-current correlation function in Eq. (\ref{cc})
into a number of vacuum correlation functions, which
are constrained by the experimental data from $e^+e^-$
annihilation, $\tau $ lepton decay, pion radiative decay,
and two-photon fusion reactions. This approach allows for
a systematic expansion in terms of pion density (and nucleon
density if baryons are included). 

\begin{figure}[htb]
\begin{center}
\epsfig{file=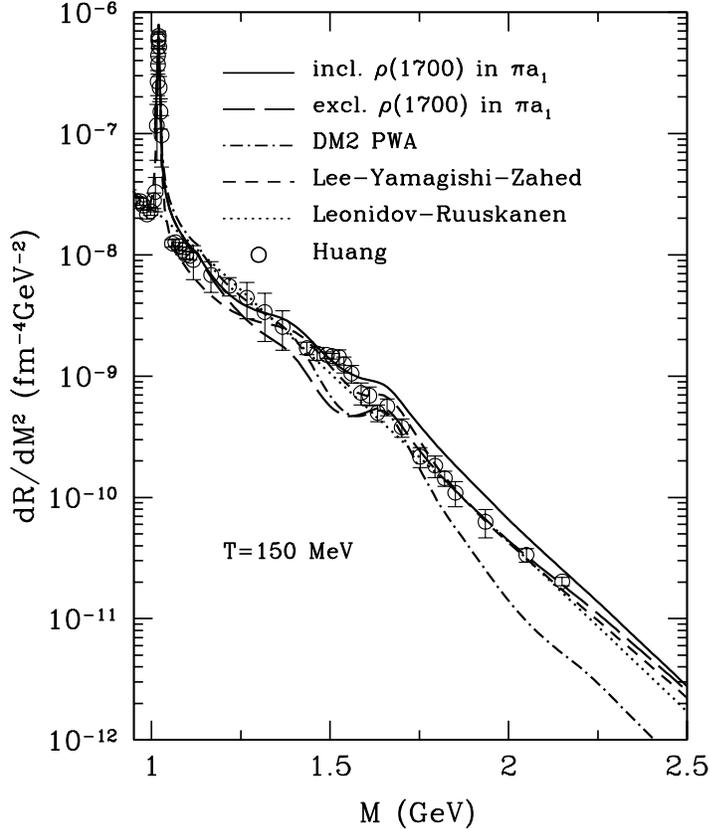,height=5in,width=5in}
\caption{Dilepton emission rates from different approaches.
\label{trtot}}
\end{center}
\end{figure}

In Fig. \ref{trtot}, we compare the dilepton emission rates
obtained with our different scenarios with those obtained in the three
models just outlined. The solid, long-dashed, and dash-dotted 
curves are our results based on our pictures for
the $\pi a_1$ cross sections. 
The scenario that includes the $\rho (1700)$ in the $\pi a_1$ form 
factor leads to a larger rate above 1.2 GeV invariant mass, as expected.
The open circles are the results of Ref. \cite{huang95} based
on empirical spectral functions. Our result with a full
$\pi a_1$ form factor is in agreement with the spectral function 
result up to 1.6 GeV invariant mass. It is slightly above at 
higher invariant masses. 
The scenario that excludes $\rho (1700)$ in the $\pi a_1$
form factor is slightly lower than the spectral function result between
1.2 and 1.6 GeV invariant mass. In other mass regions, they
give comparable rates. Finally, the use of the DM2 partial
wave analysis for the $\pi a_1$ cross section yields rates that are
somewhat lower than those obtained in the other models, passed 1.75 GeV
invariant mass. We will see later that hard background processes will
mask this difference.

The short-dashed curve gives the results of Ref. \cite{lee98}, 
which is based
on a SU(3) version of the chiral reduction formalism. The
result is in good agreement with the one from  empirical spectral
functions \cite{huang95}, suggesting  
that higher-order interactions (not explicitly included in 
Ref. \cite{huang95}) are not significant at the temperatures
considered here. Finally, the dotted curve gives the result
of Eq. (\ref{ruus}) which makes use of the hadron-quark duality
that allows the use of $R^{part}$ in the place of $R^{exp}$.
It is interesting to note that the dilepton rate from
this approach is in very good agreement with the one obtained with
empirical spectral functions
above about 1 GeV. This supports the idea that, 
even though there are resonances in the
the intermediate mass region,
these resonances cannot be easily resolved individually. 

As a partial summary, 
our results are based on kinetic theory which  assumes pair-wise
collisions of mesons. To make use of $e^+e^-$ annihilation
data as a constraint, we need to know the hadronic production
cross section in specific channels, such as $\pi\pi$, $K{\bar K}$,
or $\pi a_1$. Data are available for most of the
processes considered here. But for $\pi a_1$ an unambiguous
determination of its production cross section from the
measured $4\pi$ cross sections is not yet possible.
This introduces model dependence.
On the other hand, the other three approaches use directly
the total hadronic production cross section in $e^+e^-$
annihilation, without the need to separate them into 
specific channels and thus avoiding the uncertainties 
involved in such a separation. The nice agreement shown
in Fig. \ref{trtot} thus suggests that the model
dependence in our approach is not severe. Furthermore, this
level of agreement from different approaches for these
elementary processes provides us with the basis for 
further discussions about the dynamics of heavy-ion collisions,
the possibility of GQP formation, and charm enhancement.

\section{application to heavy-ion collisions at SPS energies}

To compare with experimental data, the thermal rate calculation is
certainly not sufficient. One needs a transport model that describes
the dynamical evolution of the colliding system, and integrate 
the dilepton production rate over the entire reaction volume and time.
In heavy-ion collisions at CERN SPS energies, 
many hadrons are produced in the initial nucleon-nucleon 
interactions. This is usually modeled by the fragmentation of 
strings, which are the chromoelectric flux-tubes 
excited from the interacting quarks. One empirically successful model 
for taking into account this nonequilibrium dynamics is RQMD 
\cite{sorge89}. To define a relativistic transport model for heavy-ion 
collisions at these energies, we have used as initial conditions the 
hadron abundances and distributions obtained from string fragmentation.  
Further interactions and decays of these hadrons are then 
taken into account as in most relativistic transport models.
This approach is found to provide a good description of 
hadronic observables (such as transverse mass spectra and rapidity 
distributions) in heavy-ion collisions at CERN SPS energies 
\cite{likob,lib97}.

We should make here some comments on the treatment of resonances and
multi-particle interactions in the transport model.
The narrow-width approximation was not used in our calculation. 
The mass distribution of the resonances (for example the $a_1$) 
in our approach is taken into account. Note that 
this realistic sampling of the spectral function is   
different from usual kinetic theory estimates of the
dilepton emission rates  where the resonance is usually given a fixed mass.
Furthermore, in our transport model, the fact that the 
resonance has a finite life time is fully respected: its
population is evolved dynamically.  The resonance is formed in 
collisions and also eventually decays. In the early 
stage of the heavy-ion collision, resonance formation is favored 
since the density is high. As the system expands and the 
density gets smaller, the resonance decay process gradually 
becomes dominant. This is again different from kinetic 
theory calculations. 
      
Let's take 
$\pi + a_1 \rightarrow dilepton$ as a concrete example. 
In our approach, a rho meson is 
first made from two-pions  colliding. It then interacts with a third 
pion to form an $a_1$ meson. Finally this $a_1$ meson annihilates 
a fourth pion to produce a lepton pair. 
The explicit treatment of 
intermediate resonances takes care of the formation time 
typical of strong interaction, which is very important for 
heavy-ion collisions at SPS energies.  Furthermore,
this algorithm also includes the possibility that
in the course of the multi-step process the intermediate
rho or $a_1$ meson might interact with other particles,
say a nucleon, rather than with the third or the fourth
pion. This amounts to a dynamical generation of a broadened
width. Finally, for any reasonable comparison with experimental data
from heavy-ion collisions, a transport model that 
describes the dynamic evolution of the colliding system
is a required tool.

\begin{figure}[htb]
\begin{center}
\epsfig{file=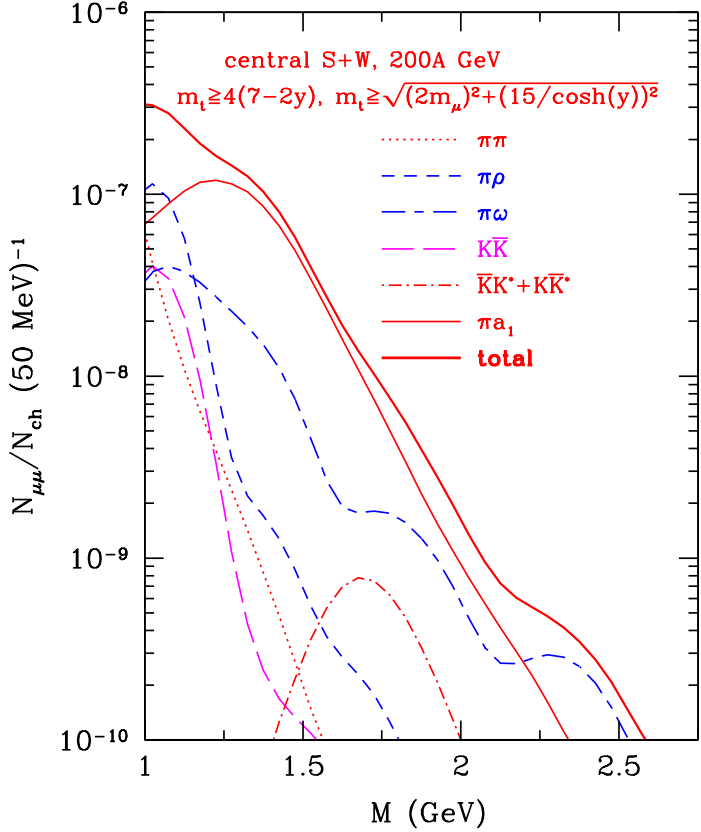,height=5in,width=5in}
\caption{Dimuon spectra in central S+W collisions at 200A GeV
from different secondary processes.
\label{swthml}}
\end{center}
\end{figure}

The contributions from the secondary processes outlined above
are shown in Fig. \ref{swthml}. These are 
obtained in our relativistic transport model \cite{likob,lib97},
including the HELIOS-3 acceptances, mass resolution, and
normalization \cite{helios}. Here $N_{\mu\mu} /N_{ch}$ 
represents number of dimuon pairs within the 50 MeV mass
bin, normalized by the charged particle multiplicity.
It is seen that the $\pi a_1$
process is by far the most important source for dimuon
yields in this mass region. The $\pi\omega$ process also
plays some role in the entire intermediate-mass region, while
the contributions from  $\pi\pi$, $\pi\rho$ and $K{\bar K}$ are 
important around 1 GeV invariant mass
The relative importance of various processes follow
basically the thermal rate predictions, as discussed in 
the previous section.

\begin{figure}[htb]
\begin{center}
\epsfig{file=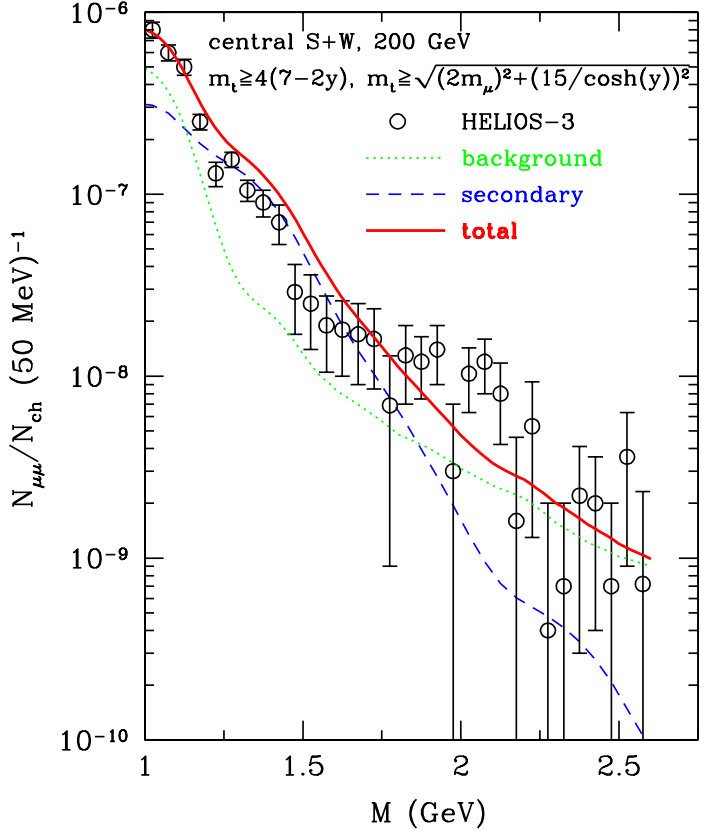,height=5in,width=5in}
\caption{Dimuon spectra in central S+W collisions at 200A GeV.
The dotted curve gives the background contribution as shown
in Fig. \protect\ref{pw}, the dashed curve gives the
contribution from secondary processes shown in Fig. 
\protect\ref{swthml}, and the solid curve gives the
sum of the two contributions.
\label{swtot}}
\end{center}
\end{figure}

In Fig. \ref{swtot}, we add the secondary contributions
obtained in our transport model to the background, and compare
again with the HELIOS-3 data for central S+W collisions. 
It is seen that the data can be adequately reproduced. This
highlights for the first time the importance of the secondary processes
for the intermediate-mass dilepton spectra in heavy-ion 
collisions. This is an important step forward in the
use of intermediate-mass dilepton spectra as a probe of
the phase transition and QGP formation. Although the current
data do not show any necessity to invoke the QGP formation
in S-induced reactions, consistent with some conclusions 
from $J/\Psi$ physics, we believe that the observation that the secondary
processes do play a significant role in the intermediate-mass
dilepton spectra is interesting and important. We note in passing that
the change of slope observed in the experimental data
corresponds in our interpretation to a crossover 
between the secondary processes and the (Drell-Yan/charm) background.
This represents a transition from soft to hard physics and this 
observation should be explored further in the future. 

\begin{figure}[thb]
\begin{center}
\epsfig{file=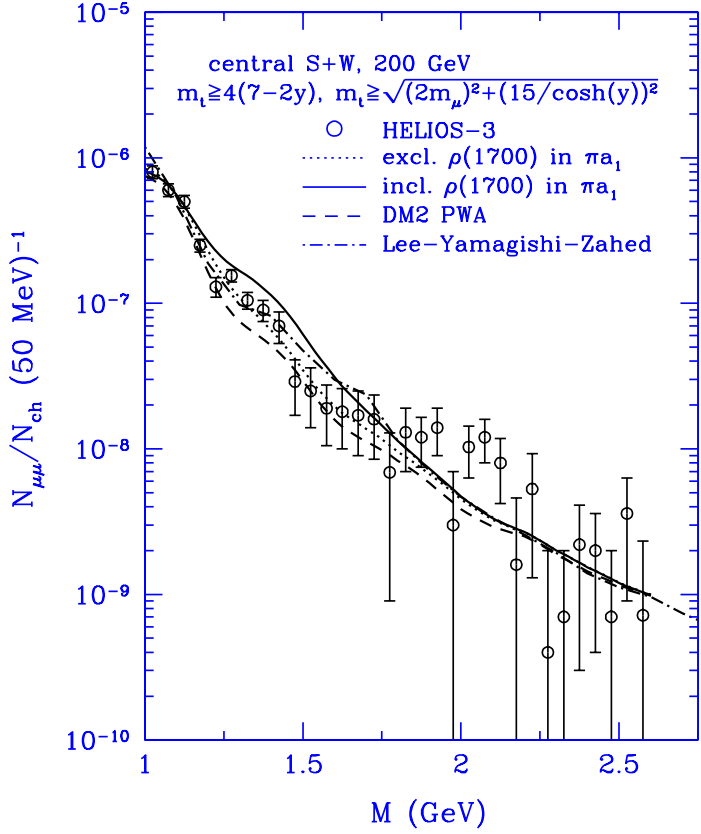,height=5in,width=5in}
\caption{Dimuon spectra in central S+W collisions at 200A GeV.
The solid and dotted curves give our results with and without
$\rho (1700)$ in the $\pi a_1$ form factor, while the
dashed curve gives our results using DM2 PWA cross section.
The dashed curve gives the results from \protect\cite{lee98}.
\label{swfm}}
\end{center}
\end{figure}

The results shown in Figs. \ref{swthml} and \ref{swtot} are
obtained under the assumption that all the 
4$\pi$ final state in the $e^+e^-$ cross section proceeds through the 
$\pi a_1$ intermediate state. We also did a calculation in which the 
$\pi a_1$ form factor contains only the normal $\rho (770)$. 
The results are shown in Fig. \ref{swfm} by the dotted curve. 
Finally, to complete our survey of possible constraints and 
uncertainties in $\pi a_1 \rightarrow e^+e^-$
cross sections, we did a calculation in which this cross section
is obtained from the $e^+e^- \rightarrow \pi a_1$ cross section
determined by the DM2 collaboration in partial wave
analysis (PWA) \cite{dm291} (see Fig. \ref{eepia1dm2}). 
The results are shown by the dashed curve in Fig. \ref{swfm}. 
The region between the solid and dashed
curves thus reflect the uncertainty for heavy-ion collisions 
due to our limited knowledge of the $\pi a_1$ cross section. This area is
not unreasonably large.  From a formal point of view, it
is fair to say that no strong evidence currently exists 
coupling the $\rho (1700)$ to a $\pi a_1$ state \cite{pdata}, even 
though better 4$\pi$ data could help resolve this issue along with 
others of interference and $\pi h_1$ contribution.
Note that the fact that the DM2 results are lower than in other
scenarios at high invariant masses is concealed by the hard background. 

Also shown in this figure is the results of Ref. \cite{lee98}
which uses the dilepton emission rate obtained in the
chiral reduction formalism with a simple fireball model for the dynamics of
heavy-ion collisions. It is see that their results are also in
good agreement with the experimental data. 

\begin{figure}[htb]
\begin{center}
\epsfig{file=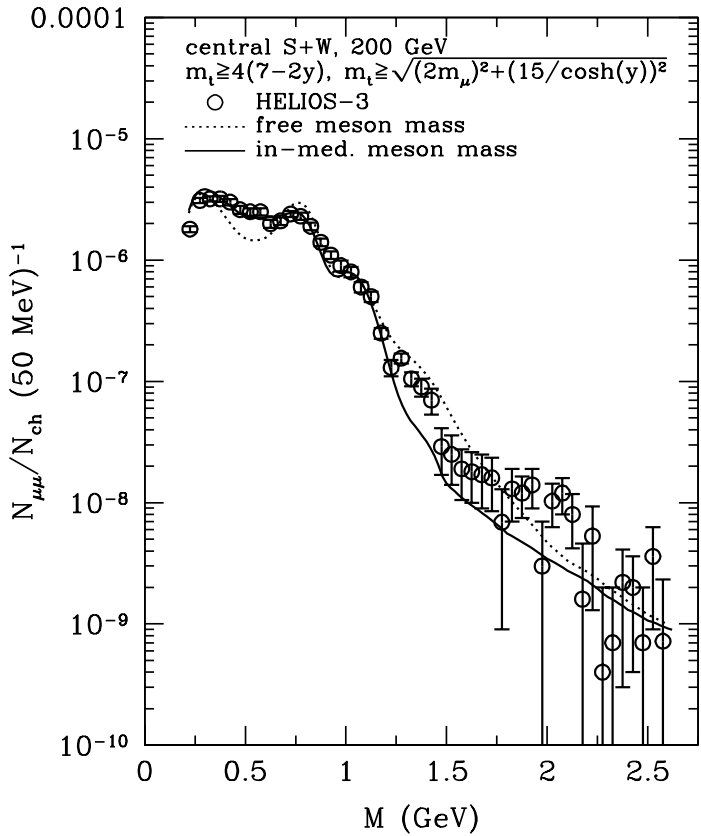,height=5in,width=5in}
\caption{Dimuon spectra in central S+W collisions at 200A GeV.
The solid and dotted curves give our results with and without
meson medium effects.
\label{swmass}}
\end{center}
\end{figure}

Another issue we want to address here is the effects of dropping vector
meson masses on the entire dimuon spectra from the threshold to 
about 2.5 GeV. In Ref. \cite{likob} it was shown that
the enhancement of low-mass dileptons could be interpreted
as a signature of vector meson masses decreasing 
with increasing density and temperature.
This should affect the dilepton spectra in the intermediate-mass
region, mainly through two effects. One is the change of the
invariant energy spectra of these secondary meson pairs. 
The second effect enters through the modification of the 
electromagnetic form factor. Since we can only conjecture how 
the masses of the higher vector resonances change with density 
and temperature, we shall assume for simplicity that they 
experience the same amount of scalar field as the ``common'' rho meson, 
namely, $m_{V,V^\prime} ^* =m_{V,V^\prime}  
-2/3g_\sigma\langle \sigma \rangle$ \cite{likob}. The results of 
this calculation are shown in Fig. \ref{swmass}.
Below 1.1 GeV and especially from 0.4 to 0.6 GeV, the agreement 
with the experimental data is much better when the dropping 
vector meson mass scenario is introduced, as was already 
shown in Ref. \cite{likob}. At higher masses, the
dropping mass scenarios somewhat underestimates the
experimental data. Also note that our starting point was our first
scenario. For completeness, however, we have to
state that there might be additional contributions from, 
e.g., secondary Drell-Yan processes \cite{spie97} that were 
not included in this study. There could also be enhanced 
production of charmed mesons. Furthermore, as we progress
higher in invariant mass, the role of baryons has to be 
carefully assessed. So far, the baryons seem to play little
role in the overall dilepton yield \cite{prak97}. This
statement was examined in Ref. \cite{steele97} for masses below 1 GeV,
which is being extended to the intermediate-mass region in \cite{lee98}.
See however Ref. \cite{ralf} for an alternate discussion on the role of
baryons. 

\section{summary and outlook}

In summary, we have carried out a study of intermediate-mass 
(between 1 and 2.5 GeV) dilepton
spectra from hadronic interactions in heavy-ion collisions.
The processes included are $\pi\pi\rightarrow l{\bar l}$,
$\pi\rho\rightarrow l{\bar l}$, $\pi a_1\rightarrow l{\bar l}$,
$\pi\omega\rightarrow l{\bar l}$, $K{\bar K}\rightarrow l{\bar l}$,
and $K{\bar K^*}+c.c \rightarrow l{\bar l}$. We calculated 
the elementary cross sections for these processes based
on chiral hadronic Lagrangians for pseudoscalar, vector,
and axial-vector mesons. The corresponding electromagnetic form
factors are determined by fitting to the experimental
data for the reverse processes of $e^+e^-\rightarrow hadrons$.
While the form factors for $\pi\pi$, $K{\bar K}$, $\pi\rho$,
$\pi\omega$, and $K{\bar K^*}+c.c.$ are uniquely
constrained by the $e^+e^-$ data, there are still
uncertainties in the $\pi a_1$ form factor, for which
we have considered three scenarios.

We used these cross sections with kinetic theory to
calculate the dilepton emission rates at finite
temperature. The $\pi a_1\rightarrow l{\bar l}$
process was found to be the most important one in the
intermediate-mass region, as was found in previous
calculations. Our rates were compared with those from other
approaches. 
This provides us with the
confidence that we have some control of the
dilepton spectra from the secondaries hadronic
interaction. This is important for the use of the
dilepton spectra as a probe of phase transition and
QGP properties.  

Finally we apply these elementary cross sections
in a relativistic transport model and calculate dilepton
spectra in S+W collisions at SPS energies. 
Again, the $\pi a_1\rightarrow l{\bar l}$ was found
to be the most important process. The comparison
of our results with experimental data from the HELIOS-3
collaboration indicates the importance of the secondary
hadronic contributions to the intermediate-mass dilepton
spectra.
 
This work can be extended in several directions. 
Although the current experimental data from the
HELIOS-3 collaboration are not very sensitive to the
different assumptions of the $\pi a_1$ form factor,
it would be very useful to pin this down 
since the $\pi a_1$ is the single most
important source of the intermediate-mass dileptons
from hadronic secondary collisions.
Also, a more elaborate calculation of the 
charmed meson production and initial (as well as possible
secondary) Drell-Yan contributions in heavy-ion
collisions is needed. Such a calculation could 
put the conclusions reached in this work concerning the
role of secondary hadronic processes and dropping
vector meson masses on an even  firmer footing. 
Finally, the current investigation can be 
extended to higher incident energies, such as those of
the RHIC collider, by combining the cross sections (or
thermal rates) obtained in this study with, e.g., 
hydrodynamical models for the evolution of heavy-ion
collisions at the RHIC energies. This kind of study is
useful for the determination of hadronic sources
in the dilepton spectra to be measured by the
PHENIX collaboration, and thus for the clear identification
of the dilepton yield arising from the QGP.  

\vskip 0.5cm

We are grateful to Nikolai Achasov, Gerry Brown, Sandy 
Donnachie, Axel Drees, Madappa Prakash, Ralf Rapp, and Ismail Zahed
for useful discussions.
This work is supported in part by the U.S. Department of Energy
under grant number DE-FG02-88ER-40388,  by the Natural Sciences
and Engineering Research Council of Canada, and by the Fonds
FCAR of the Qu\'ebec Government.

\end{document}